\documentclass[12pt]{iopart}

\usepackage{graphicx}

\newcommand{\beq}{\begin{equation}}
\newcommand{\eeq}{\end{equation}}

\newcommand{\ket}[1]{\mbox{$|#1\rangle$}}

\newcommand{\nn}{\nonumber \\}
\newcommand{\sq}[1]{\left[ {#1} \right]}

\newcommand{\ro}[1]{\left( {#1} \right)}
\newcommand{\an}[1]{\left\langle{#1}\right\rangle}
\newcommand{\st}[1]{\left|{#1}\right|}

\newcommand{\Ns}{N_{\rm S}}
\newcommand{\Nq}{N_{\rm Q}}
\newcommand{\ps}{\phi_{\rm S}}
\newcommand{\pq}{\phi_{\rm Q}}
\newcommand{\dd}{\delta\phi}

\newcommand{\text}[1]{{\rm{#1}}}

\begin{document}
\title[Heisenberg-limited phase estimation without adaptive measts]{Demonstrating Heisenberg-limited unambiguous phase estimation without adaptive measurements}

\author{B~L Higgins$^1$, D~W Berry$^{2,3, 4}$, S~D Bartlett$^5$, M~W Mitchell$^6$, H~M Wiseman$^1$ and G~J Pryde$^1$}
\address{$^1$ Centre for Quantum Dynamics, Griffith University, Brisbane, 4111, Australia}
\address{$^2$ Department of Physics, Macquarie University, Sydney, 2109, Australia}
\address{$^3$ Institute for Quantum Computing, University of Waterloo, Waterloo, ON N2L 3G1, Canada}
\address{$^4$ Centre for Quantum Computer Technology, Australia}
\address{$^5$ School of Physics, The University of Sydney, Sydney, 2006, Australia}
\address{$^6$ ICFO---Institut de Ciencies Fotoniques, Mediterranean Technology Park, 08860 Castelldefels (Barcelona), Spain}
\eads{\mailto{H.Wiseman@griffith.edu.au}, \mailto{G.Pryde@griffith.edu.au}}

\begin{abstract}
We derive, and experimentally demonstrate, an interferometric scheme for unambiguous phase estimation with precision scaling at the Heisenberg limit that does not require adaptive measurements. That is, with no prior knowledge of the phase, we can obtain an estimate of the phase with a standard deviation that is only a small constant factor larger than the minimum physically allowed value. Our scheme resolves the phase ambiguity that exists when multiple passes through a phase shift, or NOON states, are used to obtain improved phase resolution. Like a recently introduced adaptive technique [Higgins \etal 2007 \textit{Nature} \textbf{450} 393], our experiment uses multiple applications of the phase shift on single photons. By not requiring adaptive measurements, but rather using a predetermined measurement sequence, the present scheme is both conceptually simpler and significantly easier to implement. Additionally, we demonstrate a simplified adaptive scheme that also surpasses the standard quantum limit for single passes.
\end{abstract}

%03.67.-a  Quantum information (see also 42.50.Dv Quantum state
%  engineering and measurements; 42.50.Ex Optical implementations of
%  quantum information processing and transfer in quantum optics)
%42.50.St  Nonclassical interferometry, subwavelength lithography
%03.65.Ta  Foundations of quantum mechanics; measurement theory (for optical
%  tests of quantum theory, see 42.50.Xa)

\maketitle

\section{Introduction}
The interferometric measurement of an optical phase is an indispensable tool in metrology, used in length measurement, optical characterization and a myriad of other applications. Phase measurement is closely related to the more general concept of quantum parameter estimation, which is important for characterizing Hamiltonians for applications in quantum information technology, for example. In practice, there exist many different types of phase measurement problems, each with its own set of requirements and limitations, which has led to a diverse range of interferometric techniques.

In this paper, we consider the estimation of a completely unknown phase $\phi \in [0, 2\pi)$ in the situation where the total number of photon-passes through the system being probed is the crucial resource, and there is no inherent limitation to the number of times each photon can pass through the phase shift~\cite{Giovannetti2006,Boixo2006,vanDam2007,Boixo2008} (see Appendix). This corresponds to the case of a \emph{single} measurement of a \emph{fixed} phase. The number of photon-passes is important because, for example, the light may cause damage to a sensitive sample, or each sampling of the phase may be difficult to achieve (such as in probing the strength of a Hamiltonian coupling for a quantum information application~\cite{Cole2005}). In this regime, the aim is to obtain an estimate $\phi_\text{est}$ with minimal uncertainty $\Delta\phi_\text{est}$ for a given number of applications, $N$, of the phase shift~\cite{vanDam2007,Boixo2008}. In the nomenclature of quantum information, this restriction equates to having a limited number of queries of the system. 

The regime we consider may be contrasted with the ``phase sensing'' regime, such as would be found in the context of gravity-wave detection or atomic clocks. In that case the aim is to measure changes in a time-varying phase. There is an inherent limitation to the number of passes of each photon, because additional passes will eventually lead to less accurate results due to the varying phase. In addition, because the phase shifts are small, the phase is already approximately known, allowing any phase ambiguities to be resolved. In that regime, particular techniques and states of light, such as $n$-photon path-entangled ``NOON'' states~\cite{Bollinger1996,Lee2002}, are desirable due to their improved phase resolution from a single pass through the sample. The phase ambiguity in the NOON state (which only allows the phase to be estimated modulo $2\pi/n$) is resolved using the pre-existing approximate knowledge of the phase.

In phase measurement, there are two well-known limits to measurement accuracy. One is the \emph{standard quantum limit} (SQL)\footnote{In some areas this is called the shot-noise limit, with the term ``standard quantum limit'' reserved for a limit set by additional constraints which are not relevent to the situation we consider.}, which is the limit obtained by ``standard'' techniques, such as coherent states and photon counters. The other is the \emph{Heisenberg limit}, which is the fundamental limit using arbitrary quantum states and measurements. When there is an inherent limitation to the number of passes that may be used, the SQL and Heisenberg limit are $\Or(1/\sqrt{n})$ and $\Or(1/n)$, respectively, where $n$ is the number of photons~\cite{Yurke1986,Holland1993}. To attain the Heisenberg limit in this case it is necessary to create path-entangled $n$-photon states, such as NOON states~\cite{Bollinger1996,Lee2002} or other non-classical states~\cite{Yurke1986,Holland1993,Caves1981,SumPeg90,Sanders1995,BerWis00,Hofmann2006}. However, such states are very difficult to produce experimentally even for moderate $n$~\cite{Walther2004,Mitchell2004,Eisenberg2005,Leibfried2005,Sun2006,Nagata2007,Resch2007,Cable2007}.

In this paper, we follow some recent papers~\cite{Giovannetti2006,Boixo2006} in using the term Heisenberg limit, as defined above, for the general situation, regardless of whether there is a limitation on the number of passes of any individual photon. In terms of the total number of photon-passes, $N$, the Heisenberg limit is $\Or(1/N)$~\cite{Boixo2006,vanDam2007}. In order to give a meaningful basis for comparison, we take the SQL to be the maximum accuracy possible with classical states and \emph{single} passes, which is $\Or(1/\sqrt{N})$. When there is no inherent limitation to the number of passes for each photon, interference fringes with a width scaling as $\Or(1/N)$ may be obtained simply by using $N$ passes of a single photon. The difficulty is that this also yields an ambiguous phase estimate, in exactly the same way as using NOON states. When there is no additional knowledge of the phase, resolving this ambiguity is a major problem.

Recently, a number of approaches have been proposed to resolve the above ambiguity, both in the context of phase estimation~\cite{Giovannetti2006,vanDam2007} and in related contexts~\cite{Boixo2006,Rudolph2003,deBurgh2005}. These are closely related to the quantum phase estimation algorithm (QPEA) of Cleve \etal~\cite{Cleve1998,Nielsen2000} (which is the heart of Shor's factorization algorithm~\cite{Shor1994}), as well as Kitaev's phase estimation algorithm~\cite{Kitaev1996}. None of these protocols were rigorously shown, either numerically or analytically, to give Heisenberg-limited scaling for the standard deviation of the phase estimate (see \sref{sec:history}). Another scheme, not related to the QPEA, was shown numerically to be able to achieve Heisenberg-limited operation~\cite{Mitchell2005}.

Heisenberg-limited phase estimation was realized experimentally for the first time in~\cite{Higgins2007}, using a QPEA-inspired adaptive phase estimation scheme proposed in that work. That this protocol does achieve Heisenberg-limited scaling was also shown using rigorous numerics. In fact, two schemes were considered in~\cite{Higgins2007}. Firstly, the QPEA itself was implemented, showing that this scheme fails to achieve a phase uncertainty below $1/\sqrt{N}$ due to the high wings on the distribution. Secondly, a generalization of the QPEA\footnote{Note that in~\cite{Higgins2007}, the (generalized) QPEA was called the (generalized) Kitaev algorithm.} was presented and demonstrated to achieve an uncertainty within a factor of $1.6$ of the Heisenberg limit, which approaches $\pi/N$ for large $N$~\cite{SumPeg90,BerWis00,Luis,Wiseman1997}. Subsequently, a different QPEA-inspired adaptive phase estimation scheme was analytically shown to give Heisenberg-limited scaling by Boixo and Somma~\cite{Boixo2008}.

Higgins \etal~\cite{Higgins2007} demonstrated the QPEA and its generalization using multiple passes of a single photon through the phase shift. However they also noted that the same schemes could instead be implemented equivalently using single passes of NOON states. An advantage of using NOON states is that the time taken for a measurement does not scale with $N$, thus allowing the Heisenberg limit to be attained in the \emph{phase sensing} regime. A disadvantage of using NOON states, or related multiphoton entangled states, is that the size of the state scales as $N$. Another disadvantage is that the NOON-state protocols require photon-resolving detectors with inefficiency at most $\Or(1/N)$. By contrast the multipass protocols can achieve Heisenberg-limited scaling for any non-zero detection efficiency. See the end of \sref{penultimate} for further discussion.

\begin{figure}[!thb]
% expt.pdf
\center{\includegraphics[width=11cm]{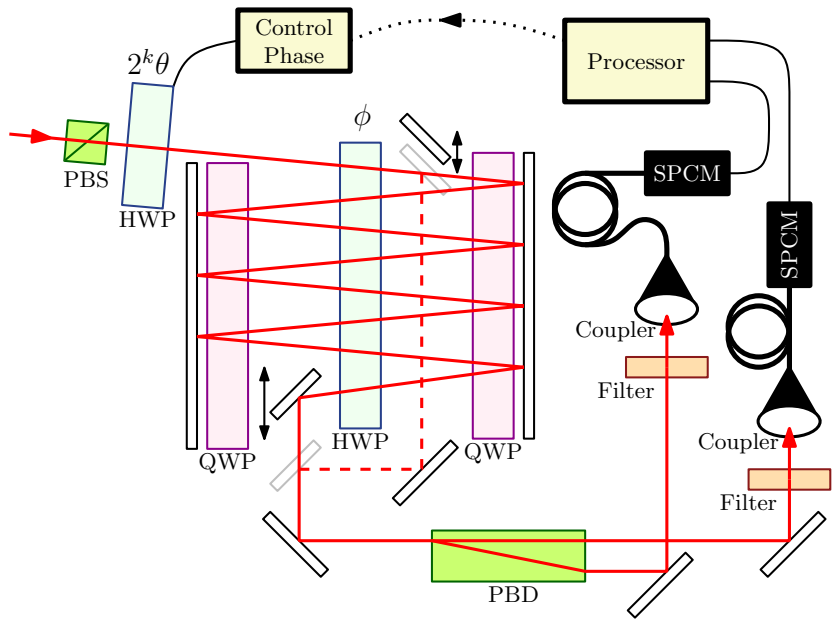}}
\caption{Schematic of the experiment. A polarizing beam splitter (PBS) creates high-fidelity equal superpositions of left- and right-circular polarization modes, which form the two arms of the interferometer. The $2^k\theta$ half-wave plate (HWP) in a motorized rotation stage implements the controlled phase shift. A photon passes multiple times through the $\phi$ HWP, implementing the multiple applications of the unknown phase shift. Quarter-wave plates (QWPs) undo $\pi$-phase shifts in the horizontal/vertical polarization basis upon reflection. Mirrors on motorized translation stages select the number of passes the photon undergoes. The photon is then discriminated by a polarizing beam displacer (PBD), filtered to remove unwanted background, and detected by a single-photon counting module (SPCM). In the hybrid scheme, measurement results determine the setting of the controlled phase shift as indicated by the dotted path from the processor to the control phase box. This path is not used for the non-adaptive scheme.}
\label{fig:expt}
\end{figure}

Here we present and experimentally demonstrate two new phase estimation schemes that offer significant simplifications compared with the generalized QPEA of~\cite{Higgins2007}, while still giving accuracy  better than $\Or(1/\sqrt{N})$. Both can be applied to either NOON states or multiple passes of single photons, as with the QPEA and its generalization. As in~\cite{Higgins2007}, we have performed the experimental demonstration using multiple passes of single photons, as shown in \fref{fig:expt}. Whereas the Heisenberg-limited scaling of the generalized QPEA of~\cite{Higgins2007} was shown numerically, we here provide analytical proofs bounding the scaling of our schemes.

The first scheme, which we shall call the \emph{QPEA--standard hybrid} or simply \emph{hybrid} scheme, combines the original QPEA with standard (single-photon, single-pass) interferometry to achieve a phase uncertainty $\Or((\ln N)^{1/4}N^{-3/4})$ or better. The second, \emph{non-adaptive} scheme uses a predetermined sequence of single-photon measurements. Remarkably, this non-adaptive scheme achieves a phase uncertainty $\Or(1/N)$, although with a higher overhead than the adaptive scheme of~\cite{Higgins2007}. As well as being of fundamental interest, the reduction in complexity offered by these new schemes makes the practical adoption of multipass phase estimation techniques more attractive.

\section{Existing schemes}\label{sec:history}
In order to compare the performance of phase estimation schemes, it is necessary to define an appropriate measure of uncertainty. Uncertainty measures are not interchangeable. For example it may be the case that although an accurate phase estimate is obtained the majority of the time, a small proportion of measurement errors are large~\cite{BerWis00}. This will lead to a large standard deviation in the phase estimate, although other measures (e.g.\ a confidence interval) may fail to capture this uncertainty. We use the standard deviation to unambiguously quantify phase uncertainty---if the standard deviation of the phase estimate scales as $\Or(1/N)$, so will other measures of uncertainty~\cite{Berry2001}. The converse is not generally true.

When considering a single measurement of an unknown phase, it is necessary for the measurement procedure to be self-contained in order to fairly evaluate it. That is, the procedure should not use additional phase information from some other source without counting the resources used to obtain that information. There are two main issues to be considered here:
\begin{enumerate}
\item NOON states or multiple passes through a phase shift provide an ambiguous phase estimate, and extra measurement steps are required to resolve the ambiguity.
\item Additional information about the phase may also be useful during the measurement procedure to design an optimal or near-optimal measurement, as is the case for measurements based on squeezed states, for example~\cite{Caves1981}.
\end{enumerate}

One approach to obtain a self-contained, unambiguous measurement is to use the QPEA~\cite{Cleve1998,Nielsen2000}. The QPEA involves two steps. In the encoding step, $K + 1$ qubits, each originally in state $(\ket{0}+\ket{1})/\sqrt{2}$, control the application of a sequence of phase shifts $\exp(\rmi  2^K\phi),\ \exp(\rmi 2^{K-1}\phi),\ldots,\exp(\rmi \phi)$. This prepares the state $\sum_{n=0}^{N}\rme^{\rmi n\phi}\ket{n}$, where $\ket{n}$ is a state in the logical basis, and $N=2^{K+1}-1$.  In the measurement step, the inverse quantum Fourier transform is used to perform a measurement in an entangled basis on the encoded state. This measurement is the optimal one for obtaining an estimate $\phi_\text{est}$ of the unknown phase $\phi$ encoded in the state~\cite{Nielsen2000,Cleve1998}. Remarkably, the measurement step can be implemented without entangling gates by using a Markovian sequence of local adaptive measurements on each of the $K+1$ qubits. At each step the measurement basis depends only on the previous measurement and its result~\cite{Griffiths1996}.

Although the measurement step is optimal for phase estimation, the encoding step is not. As a consequence, the phase uncertainty $\Delta\phi_\text{est} \equiv \sqrt{V_\text{H}}$ (where $V_\text{H}$ is the Holevo variance~\cite{Wiseman1997}) is dominated by the high wings of the distribution of phase estimates, and the QPEA does not achieve a phase estimate with uncertainty scaling at the Heisenberg limit $\pi/N$. Rather, the uncertainty actually scales as $1/\sqrt{N}$~\cite{Higgins2007}.

To achieve scaling at the Heisenberg limit, the QPEA can be modified by repeating each application of the phase shift operator $\exp(\rmi 2^k\phi)$, $k\in\{0, \ldots, K\}$, on each of $M$ independently prepared qubits~\cite{Higgins2007}. Bayesian adaptation (that is, controlling the measurement made on each qubit based on the distribution for $\phi$ conditioned on \emph{all} preceding measurements~\cite{BerWis00}) can replace the single qubit measurements to achieve $\Or(1/N)$ scaling for $M \ge 4$.

As explained above, a phase shift of $p\phi$ can be achieved by using an interferometer and either a single photon with $p$ passes through the unknown phase shift, or a NOON state with $n=p$ photons and a single pass. In the case of multiple passes, the probability distribution for the phase based on the measurement is $P(\phi|\pm) = (1\pm\sin(p\phi))/2\pi$, where the sign depends on which output the photon is detected in. In the case of NOON states, all information about $\phi$ is again contained in a binary outcome $\pm$, this time corresponding to the parity of the photon number at one of the outputs. Moreover, the probability distribution for the phase is identical to that obtained in the single-photon case. It is evident that both cases only yield information about $\phi$ modulo $2\pi/p$; as explained, the role of the QPEA, or the generalized QPEA scheme in~\cite{Higgins2007}, is to resolve this phase ambiguity. The experimental demonstration in~\cite{Higgins2007} was achieved with multiple passes.

The total resources used is quantified by the number of applications of the phase shift, and is $N=M(2^{K+1}-1)$. When using multiple passes of a photon this is equal to the total number of photon-passes, and it is equal to the total photon number for the equivalent scheme using NOON states. As discussed in~\cite{vanDam2007} and the Appendix, this is the appropriate way to quantify resources for problems of this type.

\section{QPEA--standard hybrid scheme}
For $M=1$ the Bayesian adaptive scheme of~\cite{Higgins2007} reduces to the QPEA. However, as noted above, the control protocol in the $M=1$ case has a much simpler formulation as a Markovian scheme. If the problem of outliers of $\phi_{\rm est}$ (due to the high wings of the distribution) could be solved, the QPEA would be an attractive scheme due to its simplicity.

Here we present a solution, resulting in an uncertainty that beats the SQL, in which the QPEA is supplemented by some additional simple measurements. These additional measurements are those of standard interferometry, comprising single phase shift applications on many single qubits without adaptive measurement. A related scheme was explored in~\cite{Mitchell2005}, using standard interferometry to supplement adaptive NOON state inputs with adaptive measurements; numerical simulations showed Heisenberg-limited scaling with an overhead factor of $2.03$~\cite{Mitchell2005}. We now analytically determine the scaling for the phase uncertainty using the adaptive scheme proposed above.

Consider using $\Ns$ resources to implement a standard phase measurement, in addition to the QPEA using $\Nq=2^{K+1}-1$ resources. First we consider the uncertainty in the standard phase measurement, and for simplicity alternate the controllable phase shift $\theta$ (see \fref{fig:expt}) between $\theta_1 = 0$ and $\theta_2 = \pi/2$. There are thus two sets of measurement results governed by probabilities $P_1 = (1+\cos\phi)/2$ and $P_2 = (1+\sin\phi)/2$ respectively. For even $\Ns$, we obtain estimates of $P_1$ and $P_2$, denoted $\nu_1$ and $\nu_2$, each based on $\Ns/2$ measurements. The estimate of $\phi$ is then determined from the phase of $(2\nu_1-1)+\rmi (2\nu_2-1)$. Using the Chernoff bound~\cite{Chernoff1952} (or Hoeffding's inequality~\cite{Hoeffding1963}), the probability that the errors in these estimates exceed $\epsilon$ are bounded as
\begin{equation}
P(\st{\nu_k-P_k} \ge \epsilon) \le 2\exp\ro{-2\Ns {\epsilon^2} },
\end{equation}
for $k\in\{1,2\}$. The probability that \emph{either} of the estimates differs from the $P_k$ by at least $\epsilon$ is then upper-bounded by $4\exp\ro{-2\Ns {\epsilon^2} }$. Using simple geometry, it is easily seen that if $\st{\nu_k-P_k} < \epsilon$, then the error in the angle can not exceed $\arcsin(2\sqrt 2 \epsilon)$. Let us take
\begin{equation}
\epsilon = \sqrt{3 f(\Ns) / 32 \Ns},
\end{equation}
where $f$ satisfies $f(x)\le x$ for $x\ge 1$. With this choice, we find that
\begin{eqnarray}
\arcsin \ro{2\sqrt 2 \epsilon} & = \arcsin \ro{\sqrt{3 f(\Ns) / 4\Ns}} \nn
& \le (\pi/3)\sqrt{f(\Ns)/\Ns}.
\end{eqnarray}
In the second line we have used the fact that $\Ns\ge 1$, so $2\sqrt 2 \epsilon \le \sqrt{3}/2$, and the convexity of the $\arcsin$ function over the range $[0,\sqrt{3/4}]$. Therefore, the probability of the phase error being at least $\dd=(\pi/3)\sqrt{f(\Ns)/\Ns}$ is no greater than
$4\exp\sq{-3f(\Ns)/16}$.

We now wish to place bounds on the phase uncertainty. We quantify the phase uncertainty by the square root of the Holevo variance, $V_\text{H}$~\cite{holevo84}, where
\begin{equation}
\label{eq:hol}
V_\text{H} \equiv \st{\an{ \rme^{\rmi (\phi-\phi_{\rm est})}}}^{-2} - 1.
\end{equation}
When small, the Holevo variance is well approximated by~\cite{BerWis00}
\begin{equation}
V \equiv 4\an{\sin^{2} \sq{(\phi-\phi_{\rm est})/2}}.
\end{equation}
To place an upper bound on this variance, we consider a suboptimal method of determining the phase estimate. Consider combining the standard phase estimate, $\ps$ and the QPEA phase estimate, $\pq$, by using $\pq$ when the estimates differ by less than $2\dd$, and using $\ps$ if the two estimates differ by more than $2\dd$. There are then three different alternatives which need to be considered.
\begin{enumerate}
\item
If the standard phase estimate is inaccurate, in the sense that $\st{\ps-\phi} \ge \dd$, then the contribution to $V$ can not exceed $16\exp\sq{-3f(\Ns)/16}$. Choosing $f(\Ns)=\min [\Ns, (32/3)\ln \Ns]$ ensures that this contribution is $\Or(\Ns^{-2})$.
\item
For the case where the standard phase estimate is accurate ($\st{\ps-\phi} < \dd$), if the QPEA estimate differs from the standard estimate (so $\st{\ps-\pq} > 2\dd$), then the QPEA estimate must be inaccurate (so $\st{\pq-\phi} > \dd$). Provided the QPEA uses a randomized initial setting of the controllable phase, the probability distribution is
\begin{equation}
P(\pq)=\frac{\sin^2\sq{(\Nq+1)(\pq-\phi)/2}}{2\pi(\Nq+1)\sin^2 [(\pq-\phi)/2]}.
\end{equation}
Using this probability distribution, the probability of $\st{\pq-\phi}$ being greater than $\dd$ does not exceed $[9/(\pi^2\Nq)]\sqrt{\Ns/f(\Ns)}$, and the overall contribution to $V$ then does not exceed
\begin{equation}
(1/\Nq)\sqrt{f(\Ns)/\Ns}.
\end{equation}
\item
Next consider the case that the standard phase estimate is again accurate ($\st{\ps-\phi} < \dd$) and the standard and QPEA estimates agree (so $\st{\ps-\pq} \le 2\dd$). This means that the QPEA estimate must be accurate, in the sense that $\st{\pq-\phi} < 3\dd$. With this limit, we find that the possible contribution to $V$ is bounded by
\begin{equation}
(2/\Nq)\sqrt{f(\Ns)/\Ns}+2/(\pi\Nq^2).
\end{equation}
\end{enumerate}

Collecting together all three cases gives the upper bound on $V$ as
\begin{equation}
\label{eq:hybridbound}
V \le \frac{3}{\Nq}\sqrt{\frac {f(\Ns)}\Ns}  + \Or(\Ns^{-2}) + \Or(\Nq^{-2}).
\end{equation}
Ignoring the slow variation of $f(\Ns)$, the minimum of \eref{eq:hybridbound} as a function of $N=\Nq+\Ns$ will be obtained for $\Ns=N/3$ and $\Nq=2N/3$, which gives
\begin{equation}
V=\Or(\sqrt{\ln N}N^{-3/2}).
\end{equation}

We have implemented the above hybrid scheme both experimentally (see below) and numerically. We do not consider special cases as in the above derivation, but rather determine the best estimate from the normalized product of the distributions conditioned by Bayesian updating. Also, we increment the controllable phase $\theta$ by $\pi/\Ns$ after each measurement, rather than alternating between $0$ and $\pi/2$. Numerically we find that choosing $\Ns=2^{K}$, so that $N=3\times 2^{K}-1$, is close to the optimal choice, as expected. Moreover, we find that for this choice $V_\text{H} \times N^{3/2}$ is nearly constant, also as expected (it increases from $4.83$ at $N=5$ to $6.17$ at $N=767$).

It is useful to contrast our hybrid scheme with the generalized QPEA of~\cite{Higgins2007}.  In the latter, $M$  (the number of independent iterations for each $2^{k}$-fold application of the phase shift), did not depend on $k$. In our hybrid scheme $M$ depends on $k$, as $M(K,k)=1+\delta_{k,0}2^K$. The hybrid scheme also differs in using mostly non-adaptive measurements ($2^{K}+1$ non-adaptive measurements versus $K$ adaptive ones). These observations suggest that, by considering other dependencies $M(K,k)$, good schemes may be found that dispense with adaptivity altogether.

\section{Non-adaptive scheme}
\begin{figure}[!thb]
% nonambad
\center{\includegraphics[width=10cm]{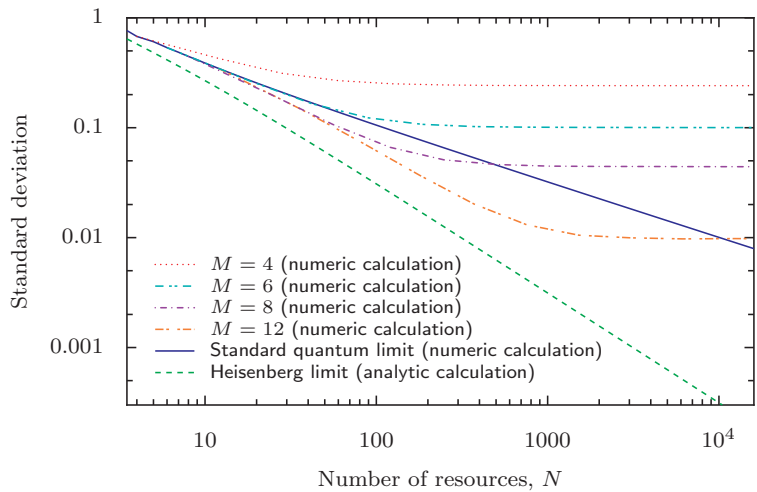}}
\caption{Predictions of standard deviations $\Delta \phi_\text{est}$ of distributions of phase estimates under a non-adaptive scheme with fixed $M$ and controllable phase $\theta$ incremented by $\pi/M$ after each measurement. Numerical predictions were based on $2^{24}$ samples. The Heisenberg limit and the SQL are shown for comparison.}
\label{fig:nonambad}
\end{figure}

In the adaptive case, it is possible to obtain scaling at the Heisenberg limit for $M\ge 4$~\cite{Higgins2007}. In contrast, for non-adaptive measurements, using small values of $M$ such as these results in much larger variances that do not decrease below some fixed amount (see \fref{fig:nonambad}). Rather than there being a threshold value of $M$ above which Heisenberg scaling is obtained, these results suggest that $M$ must increase as a function of $K$ to avoid results with large error. But this can not yield Heisenberg scaling because it leads to an increasingly large overhead.

On the other hand, as we will now show, with a suitable choice of $M(K,k)$ it is indeed possible to obtain scaling at the Heisenberg limit without adaptive measurements. Specifically, we choose $M(K,k)= M_K + \mu (K-k)$, where $M_K$ is the value of $M$ used for the $2^K$-fold application of the phase shift, and $\mu>0$. The optimal performance is obtained by numerically optimizing both of these parameters, which we consider later. First, we prove rigorously that scaling at the Heisenberg limit is guaranteed for a particular choice of $M_K$ and $\mu$.

Since our scheme is non-adaptive, it can be performed in any order; we consider using $M(K,k)$ independently prepared qubits undergoing the $2^k$-fold application of the phase shift in the sequence $k=0,1,2,\ldots,K$. Consider the results for $k=0$; applying the analysis using the Chernoff bound as in the hybrid scheme, for even $M(K,0)$ choosing $\epsilon=\sqrt{3/2}/4$ localizes the true phase to an open segment of angular size $2\pi/3$ with probability at least $1-4\exp\sq{-3M(K,0)/16}$.

Next, using this analysis on the measurements with a two-fold phase shift ($k=1$), we see that the phase is localized to two diametrically opposite segments of size $\pi/3$. As the separation of these segments is $2\pi/3$, and they are open, only one of them overlaps with the segment obtained for $k=0$. Hence, combining the results of these two sets of measurements localizes the system phase to within a segment of size $\pi/6$ with high probability. This analysis is continued up to the $2^K$-fold phase shift. At stage $k$ the phase is localized to a segment of size $(2\pi/3)/2^k$, with probability of error bounded above by $4\exp\sq{-3M(K,k)/16}$.

Consider the maximum error for each case where the phase may lie outside the segment. The maximum error in the case for $k=0$ is $\pi$, and occurs with probability no greater than $4\exp\sq{-3M(K,0)/16}$. The contribution to $\an{\sin^2\sq{(\phi_{\rm est}-\phi)/2}}$ therefore does not exceed $4\exp\sq{-3M(K,0)/16}$. Then, for $k=1,\ldots,K$, the error can not exceed $(4\pi/3)/2^k$ (for example, if the phase is outside the $k=1$ segment, the error can be as large as the size of the $k=0$ segment).  The probability of this error does not exceed $4\exp\sq{-3M(K,k)/16}$. Thus the contribution to $\an{\sin^2\sq{(\phi_{\rm est}-\phi)/2}}$ from these errors does not exceed
\begin{equation}
4\e^{-3M(K,0)/16}+4\sum_{k=1}^K \sin^2\sq{(2\pi/3)/2^k} \rme^{-3M(K,k)/16}.
\end{equation}

Finally, if the phase is correctly localized in the $k=K$ segment, the maximum error is $(2\pi/3)/2^K$.  Including this contribution we obtain the upper bound on $V$ as
\begin{eqnarray}
\label{eq:nampbound}
V \le V_\text{max}  = & 16 \rme^{-3M(K,0)/16}+(2\pi/3)^2 2^{-2K} \nn
& +16(2\pi/3)^2\sum_{k=1}^K  2^{-2k} \rme^{-3M(K,k)/16}.
\end{eqnarray}

The total number of phase shift applications is $N=\sum_{k=0}^K M(K,k) 2^k$. It is evident that, for $M(K,k)$ independent of $k$ and $K$, $V_\text{max} = \Or(1)$, in agreement with the numerical results. If we allow $M$ to be a function of $K$ but not of $k$, then the best scaling of $V_\text{max}$ is obtained for $M\propto K$, which yields $V_\text{max} = O(\ln(N)/N^2)$. To determine a variation of $M(K,k)$ that does yield scaling at the Heisenberg limit, we may take $\partial_{M(K, k)}(V_\text{max}N^2)=0$, which gives $\exp\sq{-3k\ln 2-3M(K,k)/16} = (3/\pi^2)V_\text{max}/N$.  To satisfy this, $M(K,k)$ should have the linear variation $M(K,k) = M_K+16(\ln 2)(K-k)$. With this choice, it is easy to show that
\begin{equation}
  V_\text{max} \le (2\pi/3)^2(1+32 \rme^{-3M_K/16})2^{-2K}.
\end{equation}
Since $N = \Or(2^{K})$, this proves that the phase uncertainty scales at the Heisenberg limit when we use linear variation of $M(K,k)$ with $\mu=16\ln 2$. However, $\Delta\phi_{\rm est}$ has a substantial overhead (about 54 for $M_K=23$).

The above result represents an upper bound on the phase variance. However, using numerical methods we find that  the optimal values of $\mu$ and $M_{K}$ are smaller, and yield much smaller overhead. We performed numerical simulations (up to $N=10^7$), incrementing the controllable phase $\theta$ by $\pi/M(K,k)$ after each measurement, instead of alternating between $0$ and $\pi/2$. These indicate that values of $M_K=2$ and $\mu=3$ consistently and robustly give a low overhead, less than 2.03 times the Heisenberg limit. This is only marginally above the overhead of 1.56 for the adaptive scheme ($M=6$) of~\cite{Higgins2007}, and very close to the scaling constant for the adaptive scheme of~\cite{Mitchell2005}.

\section{Experiment} \label{penultimate}
We demonstrate both the hybrid and non-adaptive schemes introduced above using photonic qubits in a common-spatial-mode polarization interferometer, as in \fref{fig:expt}, similar to that of~\cite{Higgins2007}. As with the QPEA and its generalization, these new schemes may be equivalently performed either using multiple passes of single photons or single passes of NOON states---here we use the former. The two arms of the interferometer are the right- and left-circular polarization modes, with phase shifts applied using half-wave plates. Computer-controlled motorized stages adjust the configuration of passes through the system phase and the setting of the control phase in the other arm of the interferometer. Pairs of 820~nm single photons are generated by a type-I spontaneous parametric down-conversion source, pumped by a continuous-wave laser diode. One photon of the pair is detected immediately, the other is sent through the experiment, and a coincidence detection heralds a successful measurement.

\begin{figure}[!thb]
% data.pdf
\center{\includegraphics[width=10cm]{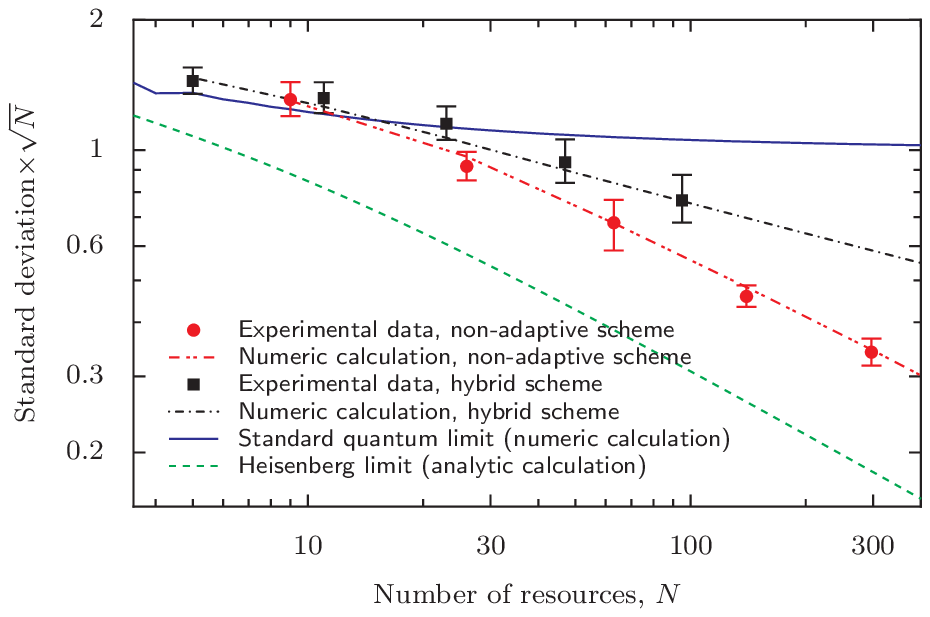}}
\caption{Performances of phase estimation schemes with varying numbers of resources, $N$. In contrast to \fref{fig:nonambad}, here we show standard deviations $\Delta \phi_\text{est}$ of distributions of phase estimates scaled by $\sqrt{N}$; in this case the SQL asymptotes to horizontal. The theoretical predictions and experimental data (each point corresponding to 1000 estimates) are compared with the SQL and the Heisenberg limit. Both the hybrid and non-adaptive schemes surpass the SQL with increasing $N$, and are consistent with respective theoretical predictions. The hybrid scheme has $\Delta \phi_\text{est} = \Or((\ln N)^{1/4}N^{-3/4})$, but the non-adaptive scheme exhibits scaling at the Heisenberg limit $\Delta \phi_\text{est} = \Or(N^{-1})$.}
\label{fig:data}
\end{figure}

The experimental data and numerical predictions for both schemes are shown in \fref{fig:data}. We vary the number of resources $N$ by varying the maximum number of passes $2^K$. The limited aperture of our optics (50~mm) lead to a limit of maximum 32 passes (maximum $K=5$). For this reason, the hybrid scheme had a maximum $N$ of 95, and the non-adaptive scheme (with $M_K=2$ and $\mu=3$) a maximum $N$ of 297.

The results for the hybrid scheme agree with our theoretical prediction of an asymptotic scaling of $\Delta \phi_\text{est}$ close to $\Or(N^{-3/4})$. This shows the interesting result that the SQL can be surpassed by combining two measurements that do not do so individually. The data for the non-adaptive scheme are also consistent with numerical predictions, clearly surpassing the SQL and exhibiting scaling at the Heisenberg limit $\Delta \phi_\text{est} = \Or(N^{-1})$. For the maximum number of resources used, the data demonstrate the predicted overhead factor of 1.91 (slightly below the asymptotic factor of 2.03) relative to the Heisenberg limit.

Finally, we make some observations about the role of photon loss in the techniques described. With either NOON states or multipass single photons, the interferometric scheme is relatively sensitive to loss in the sample~\cite{Huver2008}, with efficiency $\eta_\text{overall} \propto \eta^N$, where $\eta$ is the efficiency for a pass of a single photon through the sample. However, the multipass scheme is less sensitive to loss at the detectors, where $\eta_\text{overall} \propto \eta$, as opposed to $\eta_\text{overall} \propto \eta^N$ for NOON states. To reduce the effect of loss, it may be valuable to employ phase-sensitive states that are loss-tolerant, such as those described in~\cite{Dorner2009,Huver2008}. It should be possible to configure both adaptive~\cite{Higgins2007} and non-adaptive schemes for use with these states.

\section{Conclusions}
When using an $n$-photon NOON state, or $n$ passes of a single photon, in a two-mode interferometer with an unknown phase $\phi$ in one mode, the output probabilities vary sinusoidally with $n\phi$ rather than $\phi$. Although this gives phase resolution of $\Or(1/n)$ in principle, it does not give a phase estimate with this accuracy, because it is sensitive only to changes modulo $2\pi/n$. This ambiguity poses a fundamental problem, because the accuracy of the prior information about the phase required to resolve the ambiguity is of the same order as the phase resolution obtained by the measurement. In this paper we have devised two simple yet highly efficient techniques for eliminating the phase ambiguity. For the second technique, we obtain an unambiguous phase estimate with accuracy scaling as $\Or(1/N)$ in the total number of photon-passes $N$. This is the Heisenberg limit scaling, in that it is the best possible under the restriction of the uncertainty principle for $N$ and $\phi$. Moreover, we have experimentally demonstrated our schemes using multiple passes of single photons.

In contrast to the methods presented in~\cite{Boixo2008,Higgins2007}, the Heisenberg-limited scheme introduced here does not require adaptive measurements, but instead uses a predetermined sequence of measurements. Reducing or removing adaptive measurements should make practical implementation of precision phase estimation far easier. The crucial feature which enables $\Or(1/N)$ scaling with nonadaptive measurements is the use of larger numbers of repetitions for the measurements with fewer passes. Using this, we have analytically proven that $\Or(1/N)$ scaling is obtained. In contrast, it is not possible to obtain the Heisenberg limit scaling with nonadaptive measurements if the number of repetitions is held constant. 

\ack
This work was supported by the Australian Research Council and Spanish MEC projects QOIT (Consolider-Ingenio 2010) and ILUMA (MWM).

\appendix
\section{Quantifying resources} \label{sec:res}
In this work we quantify the resources by the total number of applications of the phase shift, regardless of whether they are applied in parallel (to different photons) or in series (repeatedly to the same photon). This allows the resources to be treated in a consistent way, yielding a theoretical limit scaling as $\Or(1/N)$~\cite{vanDam2007}. One might argue that the resources could be quantified by a pair of numbers: the number of photons and the number of passes. Then an $n$-photon NOON state with a single pass and $n$ passes of a single photon would be regarded as using different resources. However, this distinction cannot be maintained rigorously, because the experiment can be designed such that exactly the same quantity is being measured in either case. We now demonstrate this fact.

\begin{figure}[!thb]
% thought.eps
\center{\includegraphics[width=11cm]{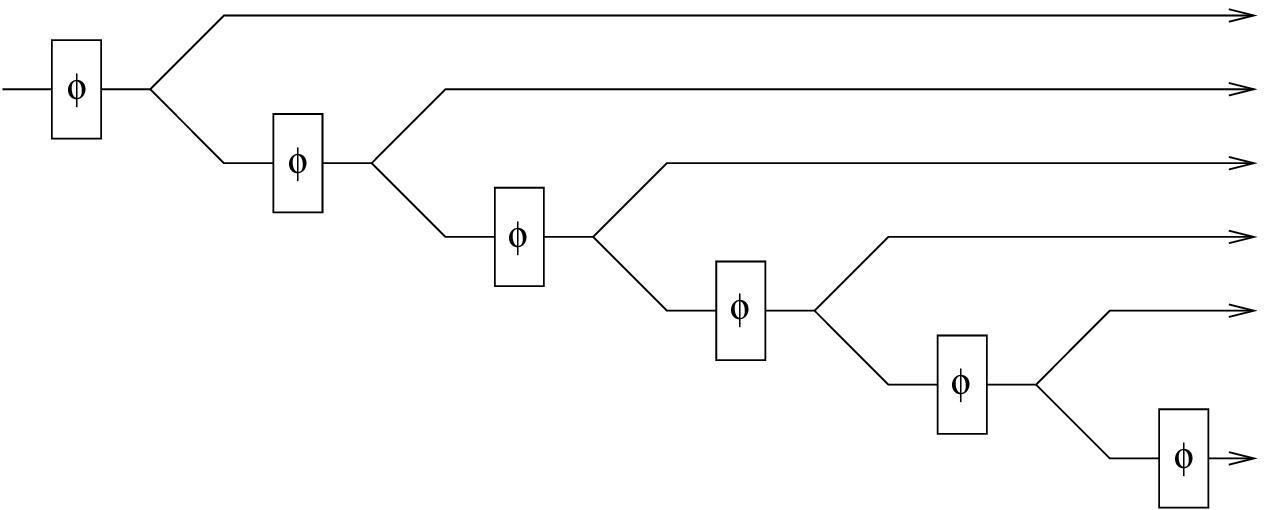}}
\caption{A thought experiment intermediate between phase measurements with a NOON state and multiple passes of a single photon. The horizontal lines are photons, and the splittings indicate that the photon has been duplicated coherently.}
\label{fig:thought}
\end{figure}

Consider the case that the NOON state is detected by time-resolving photon counters. For a two-mode state of sufficient duration, the timings of the $N$ detections will typically be distinct. The time that each photon is detected can be used to determine the time that it passed through the phase shift. (For the experiment we consider superpositions of left- and right-circular polarization, so the photon always passes through the phase shift, in contrast to the case where different spatial modes are used.) Given a set of arrival-time data $\{t_m\}$ from a NOON-state experiment, one can design a single-photon experiment (using a set of mirrors and electro-optic switches), such that the single photon passes through the sample at the same times $t_m$. In each case the phase shift is sampled in exactly the same way, by photons passing through at times $t_m$. Because the photons are indistinguishable, it makes no difference whether it is $n$ photons in a NOON state, or the same photon passing $n$ times. Similar ideas have been explored in~\cite{Giovannetti2006,vanDam2007,Boixo2008,Rudolph2003,deBurgh2005,Mitchell2005}. Interestingly, it is not difficult to devise a series of thought experiments that interpolate between the NOON state experiment and the multiple-pass experiment. For example, consider the case where the photon passes through the sample $n$ times, but it is duplicated between passes. This duplication happens in a coherent way, so the state $\alpha\ket{0}+\beta\ket{1}$ becomes $\alpha\ket{00} + \beta\ket{11}$, where 0 indicates left-circularly polarized, and 1 indicates right-circularly polarized (see \fref{fig:thought}). The interpretation is that the lower photon in the split is the ``same'' photon, so the photon passes $n$ times through the sample. On the other hand, because the photons are indistinguishable, we could equally interpret the upper photon in the split as the ``same'' photon. Then we have $n$ different photons, each passing through the phase shift once. This experiment can then be transformed to the NOON state experiment simply by moving all duplications of the photons to the beginning.

The difference between $n$ passes of a single photon, and single passes of $n$ different photons is simply due to the arbitrary interpretation of which photon is the ``same'' photon. This therefore makes it ambiguous as to how the resources should be quantified if the photons and passes are quantified as different resources. Another difficulty with that way of quantifying resources is that in a complicated measurement involving different photons with different numbers of passes, it is ambiguous as to which number of passes should be used. In contrast, the number of passes of single photons through the phase shift is always unambiguous, and has a rigorous mathematical foundation~\cite{vanDam2007}. We emphasize that, although non-classical states and multiple passes may be regarded as equivalent, they present different experimental challenges, with different advantages and disadvantages.

\end{document}